\documentstyle[preprint,eqsecnum,prd,aps,epsf,floats]{revtex}
\begin{document}

\newcommand{\be}{\begin{equation}}
\newcommand{\ee}{\end{equation}}
\newcommand{\nl}{\nonumber \\}

\newcommand{\Dv}{{\bf D}}
\newcommand{\Dt}{\tensor{D}}
\newcommand{\Dtv}{\tensor{\bf D}}
\newcommand{\sigmav}{\mbox{\boldmath $\sigma$}}

\newcommand{\bra}[1]{{\langle #1 |}}
\newcommand{\ket}[1]{{| #1 \rangle}}
\newcommand{\chidag}{\chi^\dagger}
\newcommand{\psidag}{\psi^\dagger}
\newcommand{\singS}{{}^1S_0}
\newcommand{\tripS}{{}^3S_1}
\newcommand{\singP}{{}^1P_1}
\newcommand{\tripP}[1]{{}^3P_{#1}}
\newcommand{\qbar}{\overline{q}}

\newcommand{\hc}{{\rm h.c.}}
\newcommand{\Lc}{{\cal L}}
\newcommand{\LH}{{\rm LH}}
\newcommand{\Lb}{\bar\Lambda}
\newcommand{\gam}{\gamma \gamma}
\newcommand{\lep}{\rm ee}
\newcommand{\NRQCD}{{\rm NRQCD}}
\newcommand{\ihalf}{\mbox{$\frac{i}{2}$}}
\newcommand{\ra}{\rightarrow}

 
{\tighten
\preprint{\vbox{\hbox{CALT-68-2094}
                \hbox{hep-ph/9701353} 
		\hbox{\footnotesize DOE RESEARCH AND}
		\hbox{\footnotesize DEVELOPMENT REPORT} }}
 
\title{Annihilation of S-wave quarkonia and the measurement of $\alpha_s$ \footnote{%
Work supported in part by the U.S.\ Dept.\ of Energy under Grant no.\
DE-FG03-92-ER~40701.} }
 
\author{Martin Gremm and Anton Kapustin}
 
\address{California Institute of Technology, Pasadena, CA 91125}

\maketitle 

\begin{abstract}
We analyze the relativistic corrections to annihilation rates of S-wave
quarkonia within the framework of NRQCD. We show that order $v^2$ corrections
can be expressed in terms of the heavy quark pole mass and the quarkonium mass.
 The ratio
of hadronic to radiative annihilation rates for $\eta_b$ and $\eta_c$
can therefore be predicted accurately. The contributions
 of color-octet operators to the hadronic decay rates of spin-triplet quarkonia
 are shown to be significant, even though they arise at order $v^4$
in the velocity expansion. We provide a rough estimate of the color-octet contributions and extract the value of $\alpha_s$ from the experimental data on $\Upsilon$ decays. 
\end{abstract}

}

\newpage

\section{Introduction}
Since their discovery, heavy quarkonia have been considered an important testing
 ground for quantum chromodynamics \cite{Quarkonia}. 
By now it is well established that all qualitative features
of quarkonia (e.g., a confining potential, a positronium-like spectrum,
ratios of leptonic to hadronic widths) are in agreement with what we expect
from QCD. However, in most cases we still do not have a fully quantitative 
description based on first principles. An important step towards
such a description was made in Ref.~\cite{BBL},
 where a formalism of Nonrelativistic
QCD (NRQCD) was proposed. It is based upon the observation that in a heavy
quarkonium there are several widely separated momentum scales: the
typical kinetic energy of the heavy quark, $Mv^2$, is much smaller than the
inverse size of the quarkonium, $Mv$, which in turn is much smaller than the
heavy quark mass $M$. NRQCD allows one to factor the annihilation
and production rates for quarkonia into perturbatively calculable
short-distance coefficients and nonperturbative long-distance matrix elements.
This justifies the assumption of ``naive factorization''
for S-wave quarkonium.
 On the other hand, NRQCD elucidates the role of the higher Fock components
of the quarkonium wavefunction and explains why naive
 factorization fails for P-wave states \cite{BBL}.

 NRQCD provides a
rigorous definition of long-distance matrix elements and thus allows, in 
principle, their calculation on the lattice. 
Still, given the immaturity of present day lattice simulations, one may
 ask what one can learn from quarkonia without plunging into a
 full-fledged
 lattice NRQCD computation. What we have in mind here is, first of all, a more
 accurate determination of $\alpha_s$ from the ratio of hadronic to 
electromagnetic widths \cite{Hinch}. Such a determination
 in particular could help to clarify the long-standing problem 
 of a possible discrepancy between low-energy and high energy measurements of $\alpha_s$ \cite{Shif}.

In this paper we analyze the relativistic corrections to the annihilation rates of S-wave quarkonia (both spin-singlet and spin-triplet) and apply the results
 of this analysis to restrict the
value of $\alpha_s$. The paper is organized
as follows. In section~\ref{two} we show that order $v^2$ corrections to the
color-singlet part of the annihilation rate can be expressed through the mass
of the quarkonium and the heavy quark pole mass. The hadronic widths of the
spin-triplet states, $\psi$ and $\Upsilon$, contain also a piece due to the
annihilation of the quark-antiquark pair in a color-octet state. In
section~\ref{three} we provide a rough estimate of the latter contribution
based on the running of the color-octet matrix elements. These results are
used in section~\ref{four} to extract $\alpha_s$ from the ratio of hadronic to
leptonic widths of $\Upsilon(1S)$.

\section{Relativistic corrections to S-wave quarkonium annihilation rates
in NRQCD}
\label{two}

The Lagrangian of NRQCD \cite{BBL} is
\be
\Lc_\NRQCD \;=\; \Lc_{\rm light} \;+\; \Lc_{\rm heavy} \;+\;
\delta\Lc ,
\label{LNRQCD}
\ee
where $\Lc_{\rm light}$ is the usual QCD Lagrangian for gluons and
 light quarks. $\Lc_{\rm heavy}$ is given by
\be
\Lc_{\rm heavy}
\;=\; \psidag \, \left( iD_t + \frac{\Dv^2}{2M} \right)\, \psi
\;+\; \chidag \, \left( iD_t - \frac{\Dv^2}{2M} \right)\, \chi ,
\label{Lheavy}
\ee
with $\psi$ being an operator annihilating a heavy quark, and $\chi$ being
an operator creating a heavy antiquark. Both $\psi$ and $\chi$ belong to the
fundamental representation of the color group $SU(N_c)$.
The last term, $\delta\Lc$, includes relativistic corrections to
$\Lc_{\rm heavy}$ and is of order $v^2$ compared to it. 

The annihilation of the quarkonium is a short distance process (the
 characteristic momentum scale is of order $M$) which, in the framework of 
NRQCD, is described by adding 4-quark local operators to  $\delta\Lc$. The
corresponding term has the following structure:
\be
 \delta{\Lc}_{\rm 4-fermion}
\;=\; \sum_n {f_n(\Lambda) \over M^{d_n-4}} {\cal O}_n(\Lambda) ,
\label{Lcontact}
\ee
where $d_n$ is the canonical dimension of ${\cal O}_n$.
The dimensionless coefficients $f_n(\Lambda)$ depend on the Wilsonian cutoff
 $\Lambda$ needed
to define NRQCD and can be calculated by matching the NRQCD amplitudes
generated by 4-fermion terms with annihilation contributions to the scattering
in full QCD. Following Ref.~\cite{BBL}, we take $\Lambda\sim M$.

First let us collect the expressions for the decay rates of S-wave quarkonia
including the first relativistic corrections.
According to Ref.~\cite{BBL}, the inclusive decay rates of $\eta_c$ and
$\eta_b$ to light hadrons and to two photons are given by
\begin{eqnarray}
\Gamma(\eta_{c,b} \to \LH)
&=& {2 \; {\rm Im \,} f_1(\singS) \over M^2} \;
	\bra{\eta_{c,b}} {\cal O}_1(\singS) \ket{\eta_{c,b}} \nl
&& \;+\; {2 \; {\rm Im \,} g_1(\singS) \over M^4} \;
	\bra{\eta_{c,b}} {\cal P}_1(\singS) \ket{\eta_{c,b}}
\;+\; O(v^4 \Gamma), \nl
\Gamma(\eta_{c,b} \to \gamma \gamma)
&=& {2 \; {\rm Im \,} f_{\gam}(\singS) \over M^2} \;
	 \bra{\eta_{c,b}} {\cal O}_1(\singS) \ket{\eta_{c,b}} \nl
&& \;+\; {2 \; {\rm Im \,} g_{\gam}(\singS) \over M^4} \;
	 \bra{\eta_{c,b}} {\cal P}_1(\singS) \ket{\eta_{c,b}}
\;+\; O(v^4 \Gamma),
\label{singletrates}
\end{eqnarray}
where ${\cal O}_1(\singS)=\psidag \chi \, \chidag \psi$, $
{\cal P}_1(\singS)= 1/2\left[\psidag \chi \, \chidag (-\ihalf \Dtv)^2 \psi \;+\;
 \hc\right]$.
 
For the spin-triplet S-states, $\psi$ and $\Upsilon$, the situation is more
complicated. The leading term and the order $v^2$ relativistic correction
to the $\ell^+\ell^-$ decay rate are proportional to the expectation values of
${\cal O}_1(\tripS) = \psidag \sigmav \chi \cdot \chidag \sigmav \psi$
and
${\cal P}_1(\tripS)={1\over 2}\left[\psidag \sigmav \chi
\cdot \chidag \sigmav (-\ihalf \Dtv)^2 \psi \;+\; \hc\right]$
 respectively:
\be
\Gamma(\Upsilon \to \ell^+ \ell^-)
={2 \; {\rm Im \,} f_{ee}(\tripS) \over M^2} \;
	 \bra{\Upsilon} {\cal O}_1(\tripS) \ket{\Upsilon} +
\frac{2\;{\rm Im}g_{ee}(\tripS)}{M^4}\;\bra{\Upsilon}{\cal P}_1(\tripS)\ket{\Upsilon}.
\label{Upstoee}
\ee
The decay rate of $\psi$ or $\Upsilon$ to light hadrons receives contributions
 from
 both color-singlet and color-octet components of the quarkonium
wavefunction. The color-singlet component can only decay into three or more
gluons. In contrast, the color-octet component can decay into two gluons or into
a virtual gluon which then creates a quark-antiquark pair (see Fig. 1.)
Hence, the color-octet contribution is of order $\alpha_s^2 v^4$ and may
compete with the relativistic correction to the color-singlet rate which is
of order $\alpha_s^3 v^2$. (We will see in the next section that this
 color-octet
contribution is essential for explaining experimental data on $\Upsilon$
decays.) Therefore the inclusive rate to light hadrons is
\begin{eqnarray} 
\Gamma(\Upsilon \to \LH)&=&{2 \; {\rm Im \,} f_1(\tripS) \over M^2} \;
	 \bra{\Upsilon} {\cal O}_1(\tripS) \ket{\Upsilon} +
\frac{2\;{\rm Im}g_{1}(\tripS)}{M^4}\;
	\bra{\Upsilon}{\cal P}_1(\tripS)\ket{\Upsilon}\\ \nonumber
+&&\Gamma^{(8)}(\Upsilon \to \LH).
\label{UpstoLH}
\end{eqnarray}
The color-octet part of the decay rate $\Gamma^{(8)}(\Upsilon \to \LH)$
 receives contributions from
 three four-quark operators corresponding to the three diagrams in Fig.~1:
\begin{eqnarray}
\Gamma^{(8)}(\Upsilon \to \LH) & = & {2 \; {\rm Im \,} \left(f_8(\tripP{0})+
5 f_8(\tripP{2})\right)
 \over M^4} \; \bra{\Upsilon} {\cal O}_8(\tripP{0}) \ket{\Upsilon} + \nl
 & + &  {2 \; {\rm Im \,} f_8(\singS) \over M^2} \;
\bra{\Upsilon} {\cal O}_8(\singS) \ket{\Upsilon} +
 {2 \; {\rm Im \,} f_8(\tripS) \over M^2} \;
\bra{\Upsilon} {\cal O}_8(\tripS) \ket{\Upsilon}.
\label{Gammaoctet}
\end{eqnarray}
In the latter equation we have used heavy quark spin symmetry to reexpress
the expectation value of ${\cal O}_8(\tripP{2})$ in terms of 
${\cal O}_8(\tripP{0})$.
The color-octet operators are defined as
\begin{eqnarray}
{\cal O}_8(\tripP{0})&=&{1 \over 3} \;
\psidag T^a (-\ihalf \Dtv \cdot \sigmav) \chi
	\, \chidag T^a (-\ihalf \Dtv \cdot \sigmav) \psi ,\nl*
{\cal O}_8(\singS)&=&\psidag T^a \chi \, \chidag T^a \psi ,\nl*
{\cal O}_8(\tripS)&=&\psidag \sigmav T^a \chi \cdot \chidag \sigmav T^a \psi.
\end{eqnarray}

The short-distance coefficients in the expressions for the decay rates
Eqs.~(\ref{singletrates}-\ref{Gammaoctet}) depend
on the scheme adopted to define the operators. We wish to
derive relations between the matrix elements of the operators 
${\cal P}_1(\singS),{\cal P}_1(\tripS)$ and ${\cal O}_1(\singS),
{\cal O}_1(\tripS)$, which can be used to express the decay
rates including the first relativistic corrections through the leading
order matrix elements. Since these relations also depend on the choice of the
scheme, we will discuss this issue in some detail here. 
We will limit our discussion to the operators appearing in the decay
rate of the $\eta_{b,c}$. An identical argument can be made for the operators
appearing in the decay rates of the $\Upsilon$ and $\psi$.

The vacuum saturation approximation provides the following estimate for the
matrix element of ${\cal P}_1(\singS)$

\begin{eqnarray}
\label{vacsat}
\bra{\eta_{c,b}} {\cal P}_1(\singS) \ket{\eta_{c,b}} & = &\frac{1}{2}
{\rm Re\,} \bra{\eta_{c,b}} \psidag \chi\ket{0}\bra{0}
 \left( \left(i \Dv\right)^2 \chi\right)^\dagger \psi
 + \chidag \left(i \Dv\right)^2 \psi \ket{\eta_{c,b}}.
\end{eqnarray}
The operator with two derivatives in this expression is not defined
unambiguously beyond tree level; for example, one is free to perform a shift
\begin{eqnarray}
\label{2der}
\left( \left( \left(i \Dv\right)^2 \chi\right)^\dagger \psi
 + \chidag \left(i \Dv\right)^2 \psi\right)_\Lambda &\ra& 
\left( \left( \left(i \Dv\right)^2 \chi\right)^\dagger \psi+
\chidag \left(i \Dv\right)^2 \psi\right)_\Lambda \\ \nonumber
&&+ C(\Lambda,M) \left( \chidag\psi\right)_\Lambda.
\end{eqnarray}
Here $\Lambda$ is the Wilsonian cutoff, and $C(\Lambda,M)$ is a power series
in $\alpha_s$ starting with a term of order $\alpha_s$. (In what follows we will
omit the subscript $\Lambda$, with the understanding that all operators are
regularized using the Wilsonian cutoff.) Among all possible
definitions of the operator with two derivatives, only a subset satisfies
the NRQCD velocity counting rules.
According to these rules, the matrix element of the
operator with two derivatives
should scale as ${\bf p}^2$ relative to the matrix element of $\chidag\psi$
as ${\bf p}\to 0$:
\be\label{choice}
\frac{\bra{0}\left( \left(i \Dv\right)^2 \chi\right)^\dagger \psi
 + \chidag \left(i \Dv\right)^2 \psi\ket{{\bf p},-{\bf p}}}
{\bra{0}\chidag\psi\ket{{\bf p},-{\bf p}}}={\cal O}({\bf p}^2).
\ee
Imposing this condition removes the freedom to redefine the operator as
in Eq.~(\ref{2der}).

With the convention Eq.~(\ref{choice}) it is particularly simple to determine
the short-distance
coefficient of the operator ${\cal O}_1(\singS)$ at next-to-leading order (NLO)
by comparing the $v^{-1}$ and $v^0$ contributions in full QCD and NRQCD.
On the NRQCD side it is sufficient to calculate the annihilation of a
quark-antiquark pair via the operator ${\cal O}_1(\singS)$. 
The choice Eq.~(\ref{choice}) guarantees that there are no NLO contributions
from the operator ${\cal P}_1(\singS)$ proportional to $v^{-1}$ or $v^0$.

Other ways of defining the operators are, of course, possible.
If the ${\bf p}\to 0$ limit of the expression on
 the right-hand side of Eq.~(\ref{choice}) is nonzero, 
the operator ${\cal P}_1(\singS)$
will contribute to the annihilation rate at order $v^{-1}$ and $v^0$.
In this case one needs to know the leading order short-distance coefficient
of ${\cal P}_1(\singS)$ in order to determine that of ${\cal O}_1(\singS)$
at NLO. In fact, unless one requires operators with arbitrarily many
derivatives to satisfy conditions similar to Eq.~(\ref{choice}),
they all have to be taken into account in the computation of the 
short-distance coefficient of ${\cal O}_1(\singS)$ to next-to-leading order.

In Ref.~\cite{BBL} it was assumed that
the operators with two or more derivatives need not be taken into account
when computing the NLO short-distance coefficient of ${\cal O}_1(\singS)$.
In other words, it is implicit in Ref.~\cite{BBL} that 
the scaling behavior of the operators is given by the NRQCD counting rules or,
equivalently, that Eq.~(\ref{choice}) and similar equations for operators
with more derivatives are satisfied. We adopt the same convention here.

The equations of motion for the quark fields to leading order in $v^2$ are
\be
\left( iD_t + \frac{\Dv^2}{2M} \right)\, \psi=0, \qquad
\left( iD_t - \frac{\Dv^2}{2M} \right)\, \chi=0.
\label{eqmotion}
\ee
They can be used to trade the spatial derivatives in the operator with two 
derivatives in Eq.~(\ref{vacsat}) for time derivatives acting on the quark
fields:
\begin{eqnarray}
\label{ren}
\left[\left( \left(i \Dv\right)^2 \chi\right)^\dagger \psi +
\chidag \left(i \Dv\right)^2 \psi\right]+
  A(\Lambda,M)\,\chidag \psi
 &=& 2M\,i\partial_t\left(\chidag\psi\right).
\end{eqnarray}
Here $A$ is a scheme dependent coefficient whose expansion in powers
of $\alpha_s$ starts, in general, with the term of order $\alpha_s$.
The term proportional to $\chidag\psi$ has to be included in the relation
Eq.~(\ref{ren}),
because $\chidag\psi$ mixes into the operator with two derivatives under
shifts as in Eq.~(\ref{2der}). There are corrections to Eq.~(\ref{ren}) at higher orders
in $v^2$, but for our purposes it is sufficient to take into
account only the terms shown.

Let us show that with the convention Eq.~(\ref{choice}) $A(\Lambda,M)$ is zero to all orders in $\alpha_s$. Evaluating Eq.~(\ref{ren}) between vacuum and a quark-antiquark state
$\ket{{\bf p}, -{\bf p}}$ yields:
\be\label{main}
\bra{0}\left( \left(i \Dv\right)^2 \chi\right)^\dagger \psi
 + \chidag \left(i \Dv\right)^2 \psi\ket{{\bf p},-{\bf p}}=
\left(2{\bf p}^2-A(\Lambda,M)\right)\bra{0}\chidag\psi\ket{{\bf p},-{\bf p}}.
\ee
We have used the identity
\be\label{id}
\bra{0} i\partial_t\left(\chidag\psi\right)
\ket{{\bf p},-{\bf p}}=\bra{0}\left[\chidag\psi,\,H\right]
\ket{{\bf p},-{\bf p}}=
\frac{{\bf p}^2}{M}\bra{0}\chidag\psi\ket{{\bf p},-{\bf p}},
\ee
where $H$ is the NRQCD Hamiltonian. The identity Eq.~(\ref{id}) holds to all orders in $\alpha_s$ because the asymptotic
state $\ket{{\bf p},-{\bf p}}$ is an eigenstate of $H$ with eigenvalue ${\bf p}^2/M$.
Dividing both sides of Eq.~(\ref{main}) by $\bra{0}\chidag\psi\ket{{\bf p},-{\bf p}}$,
taking the limit ${\bf p}\to 0$ and using Eq.~(\ref{choice}), one sees that
$A(\Lambda,M)=0$.

Using Eq.~(\ref{ren}) with $A=0$ yields the following relation: 
\begin{eqnarray}
\label{label}
\bra{\eta_{c,b}} {\cal P}_1(\singS) \ket{\eta_{c,b}} &=&
\frac{1}{2}{\rm Re\,} \bra{\eta_{c,b}} \psidag \chi\ket{0}\bra{0}
 \left( \left(i \Dv\right)^2 \chi\right)^\dagger \psi
 + \chidag \left(i \Dv\right)^2 \psi \ket{\eta_{c,b}}\nl
& = & M {\rm Re\,} \bra{\eta_{c,b}} \psidag \chi\ket{0}\bra{0} i\partial_t
 \left(\chidag\psi\right) \ket{\eta_{c,b}}  \left( 1+O(v^2)\right) \nl
& = & ME_{\eta_{c,b}} \bra{\eta_{c,b}} {\cal O}_1(\singS)
 \ket{\eta_{c,b}} \left( 1+O(v^2)\right).
\end{eqnarray}
Here $E_{\eta_{c,b}}$ is the energy of the quarkonium state. Since in NRQCD
the rest mass of the quarks is not included in the energy of a quarkonium
state, we can express $E_{\eta_{c,b}}$ in terms of the mass of the $\eta_{b,c}$
and the quark pole mass
\begin{equation}
\label{mass}
E_{\eta_{c,b}}=M_{\eta_{b,c}}-2M.
\end{equation}
Eqs.~(\ref{label}) and (\ref{mass}) and similar relations for the operators
in the spin-triplet decay rates allow us to express the relativistic
corrections to the 
decay rates in terms of the leading order matrix elements and the "binding
energy" of the quarkonium $M_{q\bar q}-2 M_q$. For example, the decay
rate of the $\Upsilon$ to light hadrons now takes the form
\begin{eqnarray}
\Gamma(\Upsilon \to \LH)&=&{2 \; {\rm Im \,} f_1(\tripS) \over M^2} \;
         \bra{\Upsilon} {\cal O}_1(\tripS) \ket{\Upsilon}
\left( 1+\frac{M_{\Upsilon}-2 M}{M} \frac{{\rm Im \,} g_1(\tripS)}
{{\rm Im \,} f_1(\tripS)}\right) \nl*
&+&\Gamma^{(8)}(\Upsilon \to \LH),
\end{eqnarray}
and similar expressions hold for the other decay rates
Eqs.~(\ref{singletrates}) and Eq.~(\ref{Upstoee}).

All coefficients in Eqs.~(\ref{singletrates}) and 
Eqs.~(\ref{Upstoee}-\ref{Gammaoctet}), except 
$f_8(\tripP{0}),f_8(\tripP{2})$, and $g_1(\tripS)$,
have been calculated to the necessary order in $\alpha_s$ in Ref.~\cite{BBL}.
The coefficients $f_8(\tripP{0}),f_8(\tripP{2})$ can be extracted
from Eqs.~(A9-A13) of Ref.~\cite{BBL}:
\be
$Im$ f_8(\tripP{0})={\displaystyle\frac{3\pi(N_c^2-4)}{4N_c}\alpha_s^2(M)},
\quad
$Im$ f_8(\tripP{2})={\displaystyle\frac{\pi(N_c^2-4)}{5N_c}\alpha_s^2(M).}
\ee
To extract $g_1(\tripS)$ to leading order in $\alpha_s$ we need to
compute the three-gluon annihilation rate of a free quark-antiquark pair
to order $v^2$ in their relative velocity. Fortunately, this computation
has already been performed in the context of $e^+e^-$ annihilation
 \cite{QED}. We have checked the results quoted in these papers. The 
annihilation rate of a free quark-antiquark pair in a spin triplet state 
to order $v^2$ turns out to be
\be
\Gamma\left( q\bar q(\tripS)\to 3g,v\right) = 
\Gamma\left( q\bar q(\tripS)\to 3g,0\right)\left[1-v^2\frac{19\pi^2-132}
{12\pi^2-108}+O(v^4)\right],
\label{Gamfree}
\ee
where $\Gamma\left( q\bar q(\tripS)\to 3g,v\right)$ represents
 the annihilation rate of
the quark-antiquark pair in a spin-triplet state, with $v$ denoting the
 velocity of the quark in the center of mass frame. Comparing with the
 corresponding amplitude in NRQCD, we obtain
\be
\frac{{\rm Im\,}g_1(\tripS)}{{\rm Im \,}f_1(\tripS)}
 =-\frac{19\pi^2-132}{12\pi^2-108}\left(1+O(\alpha_s)\right).
\label{g1}
\ee

One consequence of the last equation is that for the spin-triplet states the
 order $v^2$ relativistic correction to the hadronic rate is unexpectedly
 large. For the $b$-quark pole mass in the range $4.6-4.9$ GeV,
the correction to the  $\Upsilon(1S)$ decay rate
can be as large as 25\%, and still bigger for radially excited states.
 For $m_c^{pole}\simeq 1.3\;{\rm GeV}$,
the correction to the $J/\psi$ hadronic decay rate is about 150\%. Its
magnitude makes one question the usefulness of the 
nonrelativistic expansion for charmonium. For spin-singlet states, the
relativistic corrections are of the expected size.

\section{estimates of the color-octet matrix elements}
\label{three}

In order to use our expressions for phenomenological applications
estimates for the color-octet contributions are needed.
Following Ref.~\cite{BBL}, we can obtain very rough estimates by solving the
renormalization group equations for the color-octet operators. To order $v^4$
and leading order in $\alpha_s$ we find
\begin{eqnarray}
\Lambda \frac{d}{d\Lambda}\bra{\Upsilon}{\cal O}_8(\singS)\ket{\Upsilon} &=&0,
\nl
\Lambda \frac{d}{d\Lambda}\bra{\Upsilon}{\cal O}_8(\tripS)\ket{\Upsilon} &=&
 \frac{4 (N_c^2-4)\alpha_s}{N_c \pi M^2}
 \bra{\Upsilon}{\cal O}_8(\tripP{0})\ket{\Upsilon},\nl
\Lambda \frac{d}{d\Lambda}\bra{\Upsilon}{\cal O}_8(\tripP{0})\ket{\Upsilon} &=&
\frac{4 C_F\alpha_s}{81 N_c \pi}(M_\Upsilon -2M)^2
\bra{\Upsilon}{\cal O}_1(\tripS)\ket{\Upsilon},
\label{RGeqs}
\end{eqnarray}
where 
\begin{equation}
\bra{\Upsilon}\psidag\sigmav(-\ihalf\Dtv)^2\chi\chidag\sigmav(-\ihalf\Dtv)^2\psi
\ket{\Upsilon}=M^2(M_\Upsilon-2M)^2
\bra{\Upsilon}{\cal O}_1(\tripS)\ket{\Upsilon}
\end{equation}
was used in Eqs.~(\ref{RGeqs}).
We can express the matrix elements at
the factorization scale $\Lambda\sim M$  in terms of those at a low scale
$\Lambda\sim\Lambda_{\rm QCD}$ by solving Eqs.~(\ref{RGeqs}).
 The color-octet operators mix between themselves and with color-singlet
 operators. Formally, the terms coming from the mixing with color-singlets
are logarithmically enhanced. To get a rough estimate of the color-octet
matrix elements, we assume that these terms dominate. This 
yields:
\begin{eqnarray}
\bra{\Upsilon}{\cal O}_8(\tripS)\ket{\Upsilon} &\approx&
 \frac{8(N_c^2-4)C_F}{81N_C^2\pi^2}\frac{(M_\Upsilon-2M)^2}{M^2}
 \left(\frac{2\pi}{\beta_0}\ln\left(\frac{1}{\alpha_s(M)}\right)\right)^2
 \bra{\Upsilon}{\cal O}_1(\tripS)\ket{\Upsilon}, \\
\bra{\Upsilon}{\cal O}_8(\tripP{0})\ket{\Upsilon} &\approx&
 \frac{4 C_F}{81 N_c \pi}(M_\Upsilon -2M)^2\,\frac{2\pi}{\beta_0}
 \ln\left(\frac{1}{\alpha_s(M)}\right)
 \bra{\Upsilon}{\cal O}_1(\tripS)\ket{\Upsilon}.
\end{eqnarray}
In the same spirit we set
$\bra{\Upsilon}{\cal O}_8(\singS)\ket{\Upsilon}\approx 0$ since it does not 
acquire a logarithmically enhanced contribution.

In order to check whether these estimates are reasonable, we consider the following ``ratio of ratios'':
\be
R_{mn}(\Upsilon)=\frac{\Gamma\left(\Upsilon(mS)\to \LH\right)/\Gamma\left(\Upsilon(mS)\to
\ell^+\ell^-\right)}{\Gamma\left(\Upsilon(nS)\to \LH\right)/\Gamma\left(\Upsilon(nS)\to \ell^+\ell^-\right)}.
\label{ratio}
\ee
Substituting Eq.~(\ref{UpstoLH}) and a similar expression for the dileptonic
rate into Eq.~(\ref{ratio}) we get
\be
R_{mn}=1-8.0\frac{M_m-M_n}{M_m}+\frac{\Gamma^{(8)}\left(\Upsilon(mS)
\to \LH\right)}{\Gamma\left(\Upsilon(mS)
\to \LH\right)} - \frac{\Gamma^{(8)}\left(\Upsilon(nS)
\to \LH\right)}{\Gamma\left(\Upsilon(nS)\to \LH\right)}+
O(v^4,\alpha_s v^2).
\ee
Neglecting the color-octet contribution completely, we obtain $R_{12}\simeq
1.5$, in disagreement with the experimental value $R_{12}=0.95\pm 0.15$
\cite{PDG}.
In order to evaluate the color-octet contribution numerically, we need the 
pole mass of the $b$ quark. Various methods give $M$
in the range $4.6-4.9$ GeV \cite{Bmass,QCDsmr}
corresponding to the theoretical value $R_{12}\simeq 0.95-1.53$.
 For the lower quark masses our estimates shift the value of $R_{12}$ 
much closer
to the experimental number. We take this to be an indication that our
estimates give reasonable order of magnitude values for the color-octet
matrix elements.

\section{Application to the determination of $\large \alpha_{\lowercase{s}}$}
\label{four}

As we have seen in the previous section, the leading relativistic correction
to the annihilation rates of $\eta_c$ and $\eta_b$ is expressed in terms
of the $b$-quark pole mass. The latter can be extracted from the
measurement of moments of the photon spectrum in the
inclusive $b\to s\gamma$ decay \cite{ZoltanI}, from inclusive
semileptonic $b\to c$ decays \cite{Bmass}, or from sum
rules for quarkonia \cite{QCDsmr}. Therefore the ratios of hadronic
to radiative decay rates of $\eta_c$ and $\eta_b$ are ideal for
 determining $\alpha_s$. Unfortunately, these measurements are very hard (though
not impossible) to do.
\footnote[3]{There is still one problem on the theoretical
side. Next-to-leading order (NLO) perturbative corrections to these ratios are very large
\cite{BBL,Barbieri},
and one would like to know NNLO corrections to have some idea about
the convergence of the perturbation series.} 
The $\Upsilon$s are much easier to study from the experimental point of view,
but the theoretical interpretation is complicated by the presence of the 
color-octet contribution. 
We use the results of the previous section to estimate this contribution
to the hadronic decay rate. For $M=4.6-4.9$ GeV the color-octet contribution
ranges from 0 to 9\%. Therefore we take 9\% as an estimate of the error from
neglecting it. The order $v^2$ relativistic
correction is evaluated with $M$ in the same range as above. We use the renormalization scale dependence of the NLO prediction to estimate the error from perturbative NNLO corrections.

Having adopted such estimates of theoretical uncertainties, we
use the experimental value $\Gamma\left(\Upsilon(1S)\to \LH\right)/\Gamma\left(\Upsilon(1S)\to \ell^+\ell^-\right)=37.3\pm 1.0$ \cite{PDG},
to determine 
$\alpha_s(M)=0.154-0.218$. This corresponds to
\be
\alpha_s(M_Z)=0.097-0.117
\label{alpha}
\ee
at the scale $M_Z$. The higher value of $\alpha(M_Z)$ corresponds
to the lower value of the $b$-quark pole mass.
This range for $\alpha_s(M_Z)$ overlaps
with the 1 $\sigma$ confidence interval of the LEP measurement $\alpha_s(M_Z)=
0.120\pm 0.004$ \cite{Altarelli}.
The accuracy of our extraction being limited by theoretical uncertainties,
the range in Eq.~(\ref{alpha}) should not be interpreted as a 1 $\sigma$ error.
We do not quote here the values of $\alpha_s$ obtained from $\Upsilon (2S)$
decays because the theoretical uncertainties are
much larger, and also because the accuracy of data on $\Upsilon (2S)$
is worse.

Further improvements in this determination of $\alpha_s$ would come from
a more accurate
extraction of the $b$-quark pole mass, and also from a NNLO perturbative calculation
of the short-distance coefficient ${\rm Im \,}f_1(\tripS)$. For example,
knowledge of the pole mass to within 50~MeV would reduce the uncertainties
roughly by a factor of two.

We have shown that order $v^2$ relativistic corrections to annihilation
rates of the S-wave quarkonia can be expressed in terms of the quarkonium
``binding energy''. For spin-singlet 
states this observation makes it possible to predict accurately the ratio of
hadronic to radiative decay rates in terms of $\alpha_s$ and the heavy
quark pole mass. However, a calculation
of the NNLO perturbative contributions to the short-distance coefficients
is necessary to ensure that  perturbative corrections are under control.
For spin-triplet states, which are much more accessible experimentally,
the color-octet component of the quarkonium wavefunction may
contribute significantly to the hadronic annihilation rate, although the
corresponding contributions are
of order $v^4$ in the nonrelativistic expansion. Therefore, knowledge of
the expectation values of color-octet operators is needed, if we want
to predict the hadronic to leptonic ratio for spin-triplet quarkonia.
We used crude estimates based on renormalization group equations
to deduce the uncertainties due to the color-octet contributions. From the
experimental data on the $\Upsilon(1S)$ decays we extract
$\alpha_s(M_Z)=0.097-0.117$, the major part of the uncertainty coming
from the uncertainty in the $b$-quark pole mass.  
Further experimental
and theoretical efforts are needed to obtain a better estimate of $\alpha_s$
from quarkonia decays.
\acknowledgements{We thank Peter Cho, Adam Leibovich, Zoltan Ligeti, David
 Politzer and Mark Wise for helpful discussions.}

{\tighten

\newpage

\begin{figure}
\centerline{\epsfxsize=30truecm \epsfbox{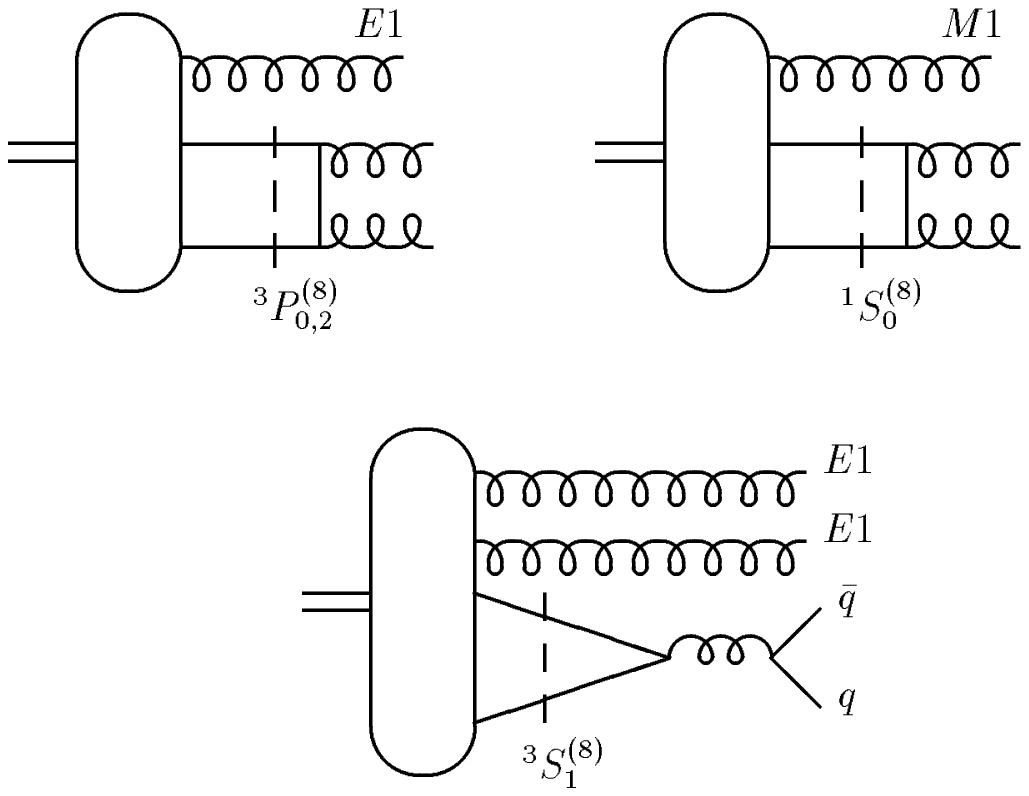} }
\caption[1]{Contributions of higher Fock states to the hadronic
 annihilation
of spin-1 S-wave quarkonia. All diagrams shown here contribute to the
rate at order $\alpha_s^2 v^4$.}

\end{figure}

\end{document}